\documentclass{jfm}
\usepackage{graphicx}
\usepackage{epstopdf, epsfig}
\usepackage{amsmath}
\usepackage[authoryear]{natbib}
\usepackage{booktabs}
\usepackage{multirow}

\shorttitle{Guidelines for authors}
\shortauthor{A.J.K.Yang}

\title{Reynolds stresses in Holmboe instabilities: from linear growth to saturation}

\author{Adam J.K. Yang\aff{1}
  \corresp{\email{jyangay@mail.ubc.ca}},
  E. W. Tedford\aff{1}
  J. Olsthoorn\aff{1}
  A. Lefauve\aff{2}
 \and G. A.  Lawrence\aff{1}}

\affiliation{\aff{1}Department of Civil Engineering, University of British Columbia,
Vancouver, BC V6T 1Z4, Canada
\aff{2}Department of Applied Mathematics and Theoretical Physics, Centre for Mathematical Sciences, University of Cambridge, Wilberforce Road, Cambridge CB3 0WA, UK}
\begin{document}

\maketitle

\begin{abstract}
The Reynolds stress in Holmboe instabilities at moderate Reynolds numbers is investigated using single wavelength simulations (SWS), multiple wavelength simulations (MWS), and laboratory experiments. The rightward and leftward propagating instabilities are separated with the two-dimensional discrete Fourier transform, enabling a direct comparison of the perturbation fields between the numerical simulations and linear stability analysis. The decomposition and superposition of the perturbation fields provide a new insight into the origin of Reynolds stresses. Conventionally, only the statistics of horizontal and vertical velocity perturbation pairs, ($u',w'$), are presented to show the degree of anisotropy in turbulent fields. Here, we present these ($u',w'$)-pairs using both theory-based and statistical approaches to reveal the mechanism of the anisotropy of perturbation field. For an individual Holmboe mode, both the simulations and linear theory show that ($u',w'$)-pairs tilt towards the 2nd and 4th quadrants ($u'w'<0$) within upper and lower vorticity interfaces, indicating an anisotropic perturbation field. This anisotropy corresponds to the tilted elliptical trajectories of particle orbits in Holmboe waves. As a result, a negative correlation between the horizontal and vertical velocity perturbation is produced, $i.e.$ negative Reynolds stresses on average. Combining the leftward and rightward Holmboe modes, ($u',w'$)-pairs are also ellipses whose orientation and aspect ratio are phase dependent. The joint probability density functions of ($u',w'$) in the linear theory and SWS show `steering wheel' structures, while in MWS and laboratory experiments the presence of waves of varying wavelength smears out the `steering wheel' structure leaving an elliptical cloud with similar orientation to the corresponding linear prediction. The vertical structure of the Reynolds stresses in the simulations and the laboratory experiment agree with the linear stability predictions.
\end{abstract}


\newpage


\section{Introduction}
Flows in the ocean and atmosphere often involve the horizontal shearing of stably stratified density layers. These shear layers are subject to hydrodynamic instabilities that cause the transition from laminar to turbulent flow, wherein irreversible mixing of the density field occurs. The best-known shear instability is the Kelvin-Helmholtz instability (KHI), which quickly grows into stationary billows and then breaks down into three-dimensional turbulence \citep[]{thorpe1971experiments, peltier2003mixing}. However, in flows where a density interface is sharper than the velocity interface, the Holmboe instability can arise \citep[]{holmboe1962behavior, browand1973laboratory}. In recent years increasing attention has been paid to the Holmboe instability and the turbulence and mixing associated with it \citep[]{koop1979instability, smyth1988finite, zhu2001holmboe, smyth2003turbulence,  tedford2009observation, tedford2009symmetric, salehipour2016turbulent, lefauve2018structure, caulfield2020layering}. 

\citet{holmboe1962behavior} analysed the instability of an idealised sharp density interface centered within a shear layer. The Holmboe instability is the result of a resonant interaction between wave-like disturbances at the edges of the shear layer and the density interface \citep[]{baines1994mechanism, carpenter2011instability}. Instability develops on each side of the density interface, such that two wave modes travel with equal and opposite phase speeds with respect to the mean flow velocity \citep[]{holmboe1962behavior, browand1973laboratory}. The first verification of the symmetric Holmboe instability, comprising two oppositely propagating Holmboe waves of equal amplitude was made through direct numerical simulations (DNS) by \citet{smyth1988finite}, subsequently \citet{zhu2001holmboe} observed symmmetric Holmboe instabilities in a laboratory exchange flow. 

Linear stability theory based on Taylor-Goldstein equation has been shown to successfully predict some of the basic properties of nonlinear Holmboe waves ($e.g.$ growth rate, wave phase speed, wavelength, and mode shape) in numerical simulations \citep[]{haigh1999symmetric, carpenter2010holmboe}, laboratory experiments \citep[]{lawrence1991stability, pouliquen1994propagating, zhu2001holmboe, hogg2003kelvin, tedford2009symmetric, lefauve2018structure}, and field observations \citep[]{tedford2009observation, smyth2011narrowband, tu2020acoustic, lian2020numerical}.

While linear theory is not formally consistent with the finite amplitude Holmboe waves investigated in the recent paper, we are motivated by the prospect that it may provide insignt into the origin of Reynolds stresses in stratified shear flows in natural geophysical systems, where the turbulence is often intermittent and spatially inhomogeneous \citep[]{sherman1978turbulence, ivey2008density, smyth2019self}.  In these flows the stratification invariably introduces a directional preference (anisotropy) in the velocity fluctuations, thereby generating Reynolds stresses \citep[]{ivey2008density, kundu2001fluid, lang2019scale}. Classical models of turbulence use statistical methods to connect mean-flow quantities with the properties of these velocity fluctuations \citep[e.g.][]{tennekes1972first, odier2009fluid, xu2012experimental, wallace2016quadrant, kundu2001fluid}. However, statistical methods do not address the physical processes underlying the anisotropy. 

The goal of the present paper is to use linear stability theory to explore the Reynolds stresses generated by Holmboe instabilities, and compare this analysis with DNS, and the laboratory experiments of \citet{lefauve2019regime}.  In \S 2, we present the governing equations for the stability to the background velocity and density profiles that are sensitive to the symmetric Holmboe instability.  In \S 3, we apply these equations to calculate the horizontal and vertical velocity fluctuations generated by counterpropagating Holmboe waves, and the resultant Reynolds stress.  Section 4 presents the Reynolds stress in DNS and examine effects of wavenumber shifting on the instabilities from linear growth to saturation.  In \S 5, we compare the Reynolds stress predicted by linear stability theory with the DNS and the laboratory experiments of \citet{lefauve2019regime}, when the flow is approximately stationary. Our concluding remarks are presented in \S 6.

\begin{figure}
	\includegraphics[scale=0.5]{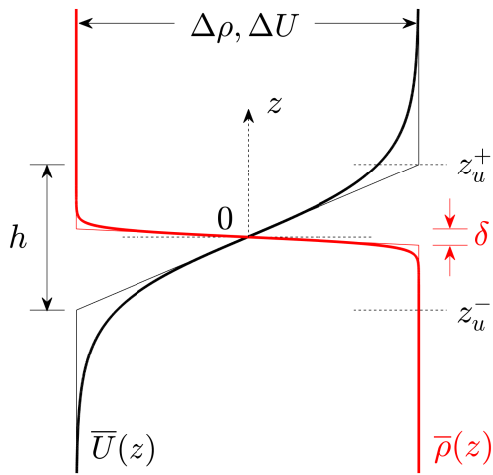}
	\centering\\
	\caption{Schematic illustrating the velocity and density profiles for $R = h/\delta =$ 9. The thick lines are smooth hyperbolic tangent profiles and the thin lines are the corresponding piecewise linear profiles. The interface locations of the upper, lower vorticities, and density are respectively $z_{u}^{+}$, $z_{u}^{-}$, and $z = $ 0.}
	\label{Schematicprofile}
\end{figure}

\section{Background}
\subsection{Setup and equations}
A density-stratified shear layer consists of initial velocity and density profiles whose variation in the vertical direction can be represented by hyperbolic tangent functions. The velocity distribution has a total jump $\Delta U$ over a length scale $h$. Similarly, the stable density distribution has a total jump $\Delta \rho$ over a length scale $\delta$:

\begin{eqnarray}\label{background}
\overline{U}(z) = \frac{\Delta U}{2} \text{tanh}(\frac{2}{h}z),  \\
\overline{\rho}(z) = -\frac{\Delta \rho}{2} \text{tanh}(\frac{2}{\delta}z).
\end{eqnarray}

A schematic illustrating these profiles is shown in figure \ref{Schematicprofile}. The ratio between the shear layer and density layer thicknesses is denoted by $R = h/\delta$. These idealised hyperbolic tangent profiles have been intensively used in literature \citep[e.g.][]{hazel1972numerical, smyth2019instability}  since they closely approximate the background profiles in stratified flows in nature. Based on these functions, the shear and density layer thicknesses respectively obtained through $(d\overline{U}/ dz)_{max}$ and $(d\overline{\rho}/ dz)_{max}$,  match the analytical results obtained using piecewise linear profile assumptions \citep[]{turner1979buoyancy, drazin2004hydrodynamic, carpenter2011instability}.   

The dimensionless Navier-Stokes equations under the Boussinesq approximation are
\begin{eqnarray}\label{eq3}
	\nabla \cdot \textbf{u} &=& 0,  \\
	\frac{\partial \textbf{u}}{\partial t}+\textbf{u}\cdot \nabla \textbf{u} &=& -\nabla p -J \rho \textbf{k}+Re^{-1}\nabla^2 \textbf{u},    \\
	\frac{\partial \rho}{\partial t}+\textbf{u}\cdot \nabla \rho &=& (Re Sc)^{-1}\nabla^2 \rho ,
\end{eqnarray} 
where $\textbf{k}$ is the unit vertical vector. We have defined three dimensionless parameters, based on the half shear layer thickness ($h$/2), half velocity scale ($\Delta U$/2) and half density scale ($\Delta \rho$/2), that characterise this system of equations: the Reynolds number ($\Rey$), the bulk Richardson number ($J$), and the Schmidt number ($Sc$) as 
\begin{eqnarray}\label{eq02}
	Re &\equiv & \frac{\frac{\Delta U}{2} \frac{h}{2}}{\nu} = \frac{\Delta U h}{4\nu},   \\
	J &\equiv & \frac{\frac{g}{\rho_0} \frac{\Delta \rho}{2} \frac{h}{2}}{(\frac{\Delta U}{2})^2} = \frac{g' h}{\Delta U^2},   \\
	Sc &\equiv & \frac{\nu}{D},  
\end{eqnarray}  
where $g'$ is the reduced gravitational acceleration. 

\subsection{Linear stability analysis}\label{sLinear}
We solve the Navier-Stokes equations described in (\ref{eq3})-(2.5) based on the assumptions of a parallel background mean flow and perturbations with normal mode forms. Similar linear stability analyses have been performed in numerous studies \citep[e.g.][]{koppel1964stability, haigh1999symmetric, carpenter2010holmboe, lefauve2018structure}. 

We consider the stability of two-dimensional perturbations. The full velocity, pressure and density fields are expressed in terms of the background field and a small superimposed perturbation ($i.e.$ $\vert u'/\overline{U} \vert \ll$ 1)
\begin{eqnarray}\label{eq5}
\textbf{u} &=& \overline{U}(z)\textbf{i}+\textbf{u}'(x,z,t),  \nonumber \\
p &=& \overline{P}(z)+p'(x,z,t),            \\
\rho &=& \overline{\rho}(z)+\rho'(x,z,t), \nonumber 
\end{eqnarray}
with all perturbations having normal mode form
\begin{equation}\label{eq6}
\psi'(x,z,t) \equiv  \mathcal{R}\{\hat{\psi}(z)\hbox{exp}(ikx+\sigma t)\}, 
\end{equation}
where $\mathcal{R}$ is taking the real part; $\hat{\psi}, \sigma \in \mathbb{C}$ and $k \in \mathbb{R}$ is the wavenumber, and $\textbf{i}$ is the unit horizontal vector. Note that in DNS, $\overline{U}$ varies slowly in time due to diffusion of the background profiles \citep{smyth1988finite}. 

Substituting into the governing equations (\ref{eq3})$-$(2.5) yields
\begin{gather}
\sigma \begin{bmatrix} \nabla^2 &  \\  & 1 \end{bmatrix} \begin{bmatrix} \hat{w}  \\  & \hat{\rho} \end{bmatrix}
=
\begin{bmatrix}
\mathcal{L}_{w} &
\mathcal{L}_{w \rho}  \\
\mathcal{L}_{\rho w} &
\mathcal{L}_{_\rho} 
\end{bmatrix}\begin{bmatrix} \hat{w}  \\  & \hat{\rho} \end{bmatrix},
\end{gather}
where
\begin{eqnarray}
\mathcal{L}_{w} &=& -ik\overline{U}\nabla^2+ik\frac{\partial^2 \overline{U}}{\partial z^2}+Re^{-1}\nabla^{4}, \nonumber \\
\mathcal{L}_{\rho} &=& -ik\overline{U}+(Re Sc)^{-1} \nabla^2, \nonumber \\
\mathcal{L}_{w \rho} &=& J(\frac{\partial^2 }{\partial z^2}-\nabla^2),  \\
\mathcal{L}_{\rho w} &=& -\frac{\partial \overline{\rho}}{\partial z}, \nonumber
\end{eqnarray}
and $\nabla^2 = -k^2+\partial^2/\partial z^2$, $\nabla^4 = k^4+\partial^4/\partial z^4-2k^2+\partial^2/\partial z^2$. The streamwise velocity eigenfunction is then reduced to $\hat{u} = (i/k) \partial \hat{w}/\partial z$. The eigenvalue can be decomposed as $\sigma = \sigma_r + i\sigma_i$, where $\sigma_r \in \mathbb{R}$ represents the growth rate of the instability and $\sigma_i \in \mathbb{R}$ is related to the phase speed $c_p = -\sigma_i/k$. No-slip and no-flux boundary conditions are imposed at upper and lower boundaries.

\section{Linear stability predictions}
In this section, we use linear stability theory to examine the particle orbits and Reynolds stresses associated with counterpropagating Holmboe waves. We highlight that the total Reynolds stress is equal to the sum of phase-independent contributions by the rightward and leftward propagating waves, and a phase-dependent interaction between them.

\subsection{Particle orbits of Holmboe waves}
Following the normal mode assumption, the velocity perturbation for a given mode is expressed as

\begin{equation}
u'(x,z,t) \equiv  \frac{1}{2}[\hat{u}(z)e^{i(kx-wt)}+c.c.], 
\end{equation}
and 
\begin{equation}
w'(x,z,t) \equiv  \frac{1}{2}[\hat{w}(z)e^{i(kx-wt)}+c.c.], 
\end{equation}
where $\omega$ ($=i\sigma$) is the frequency and $c.c.$ denotes the complex conjugate. 

The corresponding movement of fluid particles is then given by: 
\begin{subeqnarray}\label{orbiteq}
	\frac{dx_p (t)}{dt} & = &
	u'(x_p,z_p,t),\\[3pt]
	\frac{dz_p (t)}{dt} & = &
	w'(x_p,z_p,t),
\end{subeqnarray}
where $\textbf{x}(t) = x_p(t) \textbf{i}+z_p(t)\textbf{k}$.

\begin{figure}
	\includegraphics[scale=0.65]{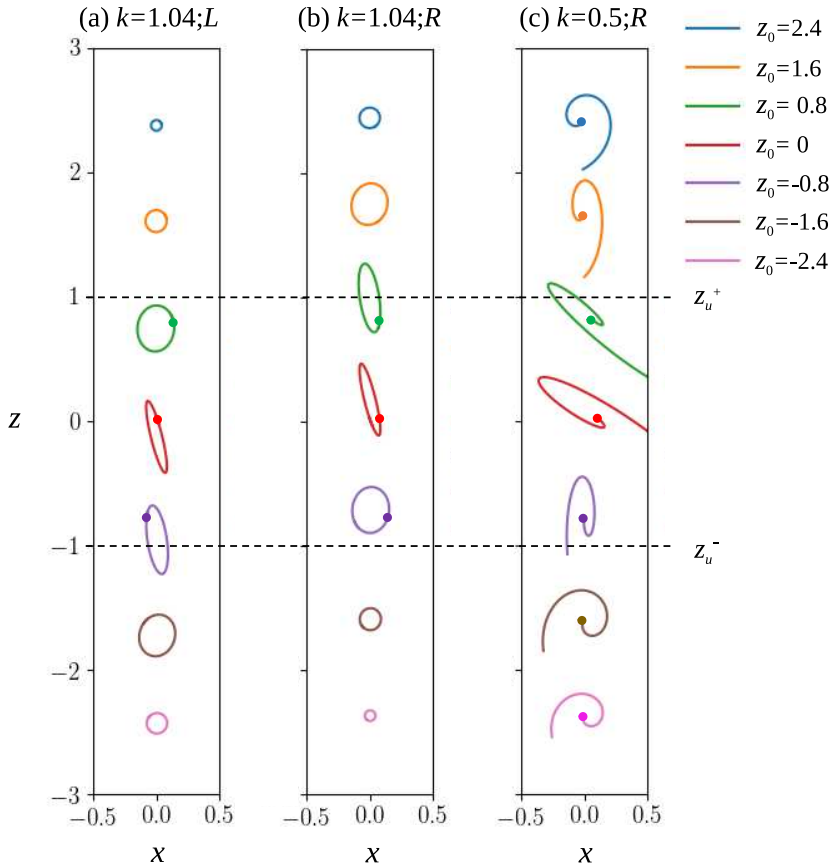}
	\centering\\
	\caption{Fluid particle orbits caused by Holmboe waves in linear stability analysis. The background profiles for the linear stability analysis is the same as the setup in DNS in table \ref{tab:kd}. The Holmboe wave is unstable for $0.01 < k < $ 1.04 (most unstable for $k = 0.5$). At $k =$ 1.04, either circular or elliptical particle orbits are generated for (\textit{a}) leftward and (\textit{b}) rightward propagating stable waves. The particle orbits for the rightward propagating waves mirror those of the leftward propagating waves. At $k = $ 0.5, the particle orbits are spirals for the rightward propagating unstable waves in (\textit{c}). $L$ and $R$ represent the leftward and rightward propagating waves, respectively. The interface locations of the upper, lower vorticities, and density are respectively $z_{u}^{+}$, $z_{u}^{-}$, and $z = $ 0.}
	\label{Orbits}
\end{figure}

Figure \ref{Orbits} shows the orbits of fluid particles (pathlines) for the stable and unstable Holmboe waves. These orbits are obtained from the time integration of (\ref{orbiteq}) at different initial vertical levels, $z_{p0}$. The centroid of $x_p$ at all levels is set at $x = 0$ for the comparison vertically. The background profile for the linear stability analysis is the same as that used in DNS (see table \ref{tab:kd}). As in the classical stability diagram \citep[e.g.][]{holmboe1962behavior}, a positive growth rate for the unstable Holmboe wave exists within a certain wavenumber range, $0.01 < k < 1.04$, with $k = 0.5$ being the most unstable wavenumber. $k = 0.01 $ and $k = 1.04$ are the wavenumbers of marginal instability. 

The stable waves exhibit closed particle orbits whereas the unstable waves exhibit open particle orbits. The particle orbits of both stable and unstable waves are often tilted towards the 2nd and 4th quadrants (figure  \ref{Orbits}), as are the corresponding ($u',w'$)-pairs. The degree of tilt depends upon the vertical location. For leftward propagating waves, the tilt is prominent within the shear layer below the density interface, $z_{u}^{-} \lesssim z \lesssim 0$ (figure \ref{Orbits}a); whereas for rightward propagating waves, the tilt is prominent within the shear layer above the density interface, $0 \lesssim z \lesssim z_{u}^{+}$ (figure \ref{Orbits}b \& c). This tilting, showing anisotropic perturbation fields, generates Reynolds stresses $\overline{u'w'}$, and will be investigated using both linear stability analysis and DNS.

\begin{figure}
	\includegraphics[scale=1]{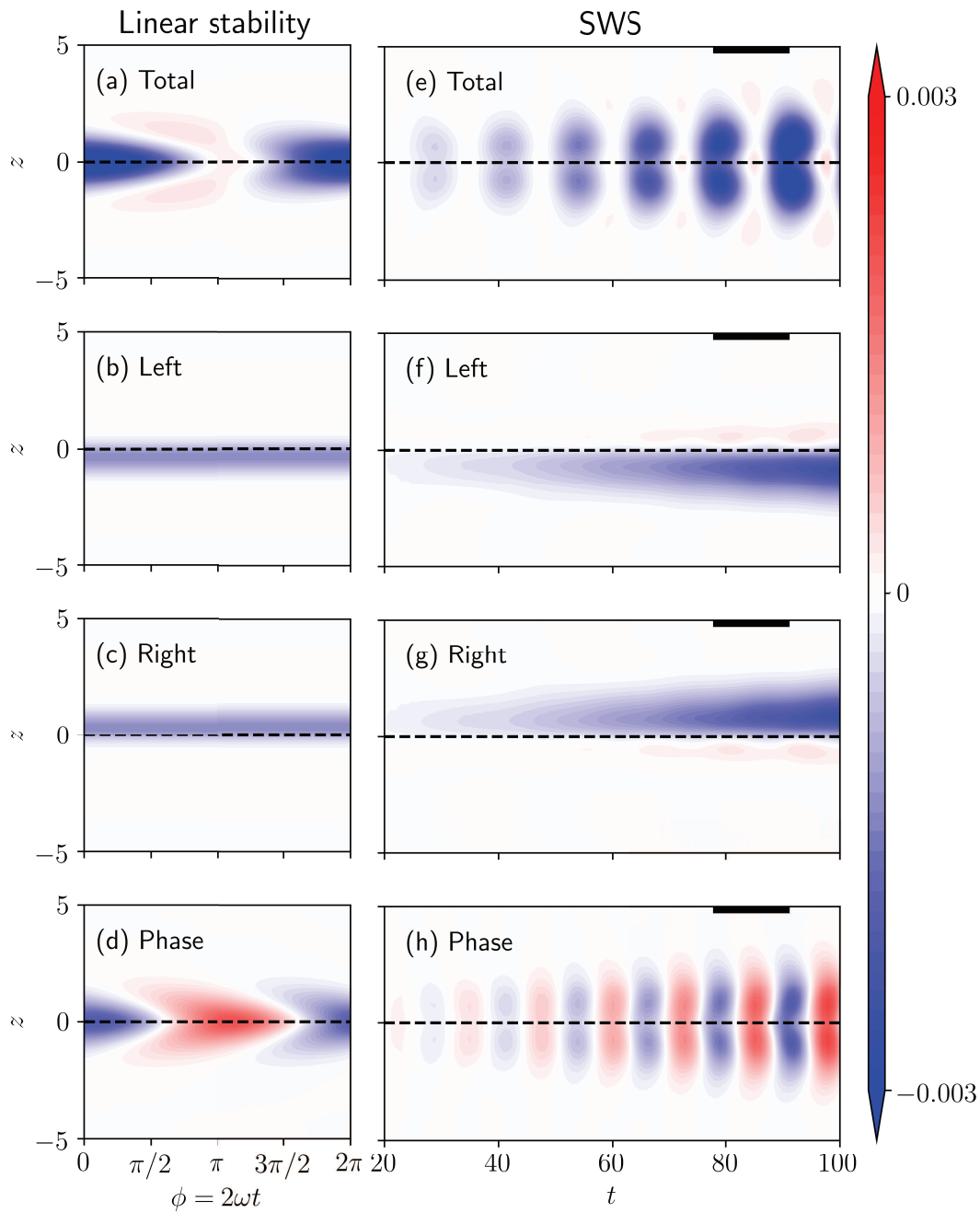}
	\centering\\
	\caption{Reynolds stress ($\langle u'w' \rangle_x$) terms in (\ref{PhasetermSimplifiedfinal}) within one wave period from linear stability analysis (the left column) and $\langle u'w' \rangle_x$ terms in nonlinear SWS (the right column). (\textit{a},\textit{e}) the total field, (\textit{b},\textit{f}) the leftward propagating wave field, (\textit{c},\textit{g}) the rightward propagating wave field, and  (\textit{d},\textit{h}) the `Phase Dependent' term. The background profile for the linear stability analysis is the same as that in SWS in table \ref{tab:kd}. In SWS, the period under the black solid line was compared with linear stability analysis in figure \ref{Com_singlebothmodes}.}
	\label{Linearanalysis}
\end{figure}

\subsection{Interaction of counterpropagating Holmboe waves}
To determine Reynolds stresses, using linear stability theory, we examine the interaction of the rightward and leftward propagating Holmboe modes: 

\begin{equation}
u' = u'^{(L)} + u'^{(R)}, ~ w' = w'^{(L)} + w'^{(R)},
\end{equation}
where the superscripts $(L)$ and $(R)$ are the components corresponding to the leftward and rightward propagating modes, respectively. 

The total velocity perturbations can be rewritten as
\begin{equation}
u' =  \frac{1}{2} [\hat{u}^{(L)}(z)e^{i(k_Lx-\omega_L t)}+ c.c.]+ \frac{1}{2} [\hat{u}^{(R)}(z)e^{i(k_Rx-\omega_R t)}+ c.c.],
\end{equation}
and $w'$ has the same form; $k_L$ ($k_R$) and $\omega_L$ ($\omega_R$) are the wave number and frequency for the leftward (rightward) propagating waves. 

To focus on the interaction between the two waves independent of their growth, we set growth rates to zero ($\omega_L, \omega_R \in \mathbb{R}$). This is consistent with statistical stationarity as commonly assumed in studies of stratified shear flow \citep[e.g.][]{osborn1980estimates, shih2005parameterization, ivey2008density, portwood2019asymptotic}. 

Then, we have $k_L = k_R = k$ and $\omega_L = -\omega_R = \omega$ for the symmetric Holmboe waves. The total horizontally averaged Reynolds stress over one wavelength is then expressed as 

\begin{eqnarray}\label{Phaseterm}
\langle u'w' \rangle_x &=& \frac{1}{4}[\hat{u}^{(L)}\hat{w}^{*(L)}+\hat{u}^{*(L)}\hat{w}^{(L)}+ \hat{u}^{(R)}\hat{w}^{*(R)}+\hat{u}^{*(R)}\hat{w}^{(R)}]      \\
&+& \frac{1}{4}\mathcal{R}\{ \underbrace{(\hat{u}^{*(L)}\hat{w}^{(R)}+ \hat{u}^{(R)}\hat{w}^{*(L)})e^{2i\omega t}}_{\text{Phase Dependent 1}} +\underbrace{(\hat{u}^{(L)}\hat{w}^{*(R)}+\hat{u}^{*(R)}\hat{w}^{(L)})e^{-2i\omega t}}_{\text{Phase Dependent 2}}\}.  \nonumber
\end{eqnarray}
One may write $\hat{u} = \hat{u}_r + i\hat{u}_i$ and $\hat{w} = \hat{w}_r + i\hat{w}_i$ ($\hat{u}_r, \hat{u}_i, \hat{w}_r, \hat{w}_i \in \mathbb{R}$), then `Phase Dependent 1' and `Phase Dependent 2' terms are a pair of complex conjugates, and thus we have 

\begin{eqnarray}\label{PhasetermSimplifiedfinal}
\underbrace{ \langle u'w' \rangle_x}_{\text{Total}} &=& \underbrace{\frac{1}{4} \hat{u}^{(L)}\hat{w}^{*(L)}+\hat{u}^{*(L)}\hat{w}^{(L)}}_{\text{Left},~ \langle u'w' \rangle_{x;L}} + \underbrace{\frac{1}{4} \hat{u}^{(R)}\hat{w}^{*(R)}+\hat{u}^{*(R)}\hat{w}^{(R)}}_{\text{Right}, ~\langle u'w' \rangle_{x;R}} \\
&+&\underbrace{\frac{1}{2}\mathcal{R}\{(\hat{u}^{*(L)}\hat{w}^{(R)}+ \hat{u}^{(R)}\hat{w}^{*(L)})e^{i \phi}}_{\text{Phase Dependent}}\},   \nonumber
\end{eqnarray}
where $\hat{u}^*$ ($\hat{w}^*$) is the complex conjugate of $\hat{u}$ ($\hat{w}$) and $\langle \cdot \rangle_i$ represents an average in the direction $i$; the time dependent variable is $\phi = 2 \omega t$. 

Here the first line on the right-hand side of (\ref{PhasetermSimplifiedfinal}) is identical to the horizontally averaged Reynolds stress for the leftward propagating and rightward propagating modes; while the second line is an additional phase-dependent term generated from the superposition of the leftward and rightward propagating waves. 

For an individual mode, the Reynolds stress does not have a phase dependent term. Averaging in the $x$ and/or $t$ over a wave cycle for the leftward propagating wave yields
\begin{equation}\label{singlemodeL}
\langle u'w' \rangle_{x;L} = \langle u'w' \rangle_{t;L} = \langle u'w' \rangle_{xt;L} = \frac{1}{4}(\hat{u}^{(L)}\hat{w}^{*(L)}+\hat{u}^{*(L)}\hat{w}^{(L)}),
\end{equation}
and the averaged Reynolds stress for the rightward propagating wave has the same form. All products are independent of $x$ and $t$.

Further taking an average of (\ref{PhasetermSimplifiedfinal}) over one wave period gives us 
\begin{equation}\label{linearsum}
\langle u'w' \rangle_{xt} = \langle u'w' \rangle_{xt;L} + \langle u'w' \rangle_{xt;R}. 
\end{equation}

Figure \ref{Linearanalysis}(a)-(d) shows the evolution of different terms in (\ref{PhasetermSimplifiedfinal}) of the Reynolds stress throughout a cycle given by linear stability analysis. In panel (a), although the perturbations spend an equal amount of time in the growth and decay portions of the cycle, the amplitude of $\langle u'w' \rangle_x$ is largest when it is negative. Thus, a net negative total Reynolds stress $\langle u'w' \rangle_{xt}$ is produced. As demonstrated theoretically, this Reynolds stress $\langle u'w' \rangle_{xt}$ is a sum of time-independent $\langle u'w' \rangle_{x;L}$ and $\langle u'w' \rangle_{x;R}$ as shown in panels (b) and (c) respectively. The Reynolds stresses for the rightward and leftward propagating waves are concentrated above and below the density interface respectively, and they are independent of time. 

The additional phase dependent term is shown in figure \ref{Linearanalysis}(d). Its cycle illustrates that the magnitude is equally distributed between decay period ($0 \leq \phi < \pi$) and the growth period ($\pi \leq \phi < 2\pi$). Over one period, its net contribution to the Reynolds stress is zero, resulting an isotropic perturbation field. It should be noted that the maximum value of the phase dependent term is larger than the magnitude of the time averaged $\langle u'w' \rangle_{xt}$, which results in $\langle u'w' \rangle_{x}$ being positive at some instants ($e.g.~\phi = \pi$). These oscillating patterns are also observed in numerical simulations. We will compare the linear stability analysis with the DNS results in \S 4.1.2.

\section{Numerical simulations}
We perform two-dimensional direct numerical simulations using Dedalus, a parallelised pseudospectral solver \citep{burns2020dedalus}, to solve the Navier-Stokes equations (\ref{eq3})$-$(2.5). The computational domain height is 20, which prevents the boundaries from interfering with the shear layer during the linear development \citep{haigh1999symmetric}. Free slip and no flux boundary conditions are imposed at $z = \pm $ 10. The horizontal length of the domain is $L_x = \lambda_0$ (the wavelength of the maximum growth rate) in single wavelength simulations (SWS), and is $L_x = 16\lambda_0$ in multiple wavelength simulations (MWS). Periodic boundary conditions were imposed in the horizontal, and we used Chebyshev grid in the vertical. A fourth-order Runga-Kutta time-stepping scheme was used. $N_x$ and $N_z$ are respectively the number of grid points in the horizontal and vertical directions. Double the number of points produce the same results. The initial ratio of the shear layer to the density layer thicknesses was $R = 9$. In this configuration, Holmboe waves appear as the dominant shear instability. 

To provide an optimal comparison with the linear stability theory, the SWS was run with an eigenfunction perturbation \citep{smyth1988finite}, which was associated with the wavenumber of maximum growth rate. The perturbation, the superposition of the leftward and rightward modes, obtained through linear stability described above, with an amplitude of $0.05 \Delta u$. Random noise with an amplitude of $\pm 0.005 \Delta u$ was also added. This perturbation method is identical to that of \citet{carpenter2007evolution}, triggering the rapid growth of Holmboe waves.

The MWS was initialised with random noise in the velocity field to stimulate the growth of the instabilities and allow different wavenumbers to evolve initially. The amplitude of the perturbation is uniformly distributed in the range $\pm 0.05 \Delta u$. A simple sinusoidal perturbation was also added to the density interface  with a wavenumber of $k_{max}$ and an amplitude of $0.05 \Delta u$. A summary of the simulations is shown in table 1.

\begin{table}
	\begin{center}
		\def~{\hphantom{0}}
		\begin{tabular}{lccccccccc}
 Run  &$N_x$ &$N_z$ &$k_{max}$ &$L_x/\lambda_0$  & Steady period & Figure \\ [3pt]

 SWS   & 128   & 1024  & 0.5   & 1  &130-210 & 3-5,7-9\\

 MWS   & 2048  & 1024  & 0.5 & 16  &250-330 & 6,7,9\\ [4pt]

		\end{tabular}
		\caption{The input parameters for the numerical simulations. The number of grid points in each direction are $N_x$ and $N_z$. The initial $\Rey = 30$, $J = 0.13$, $Sc = 256$ and $R = 9$ were used. $k_{max}$ is the wavenumber of the initial maximum growth rate and $\lambda_0 = 2 \pi/k_{max}$. As the background flow evolves over time, the nondimensional numbers, $\Rey$, $J$, and $R$ vary accordingly.}
		\label{tab:kd}
	\end{center}
\end{table}

\begin{figure}
	\includegraphics[scale=0.75]{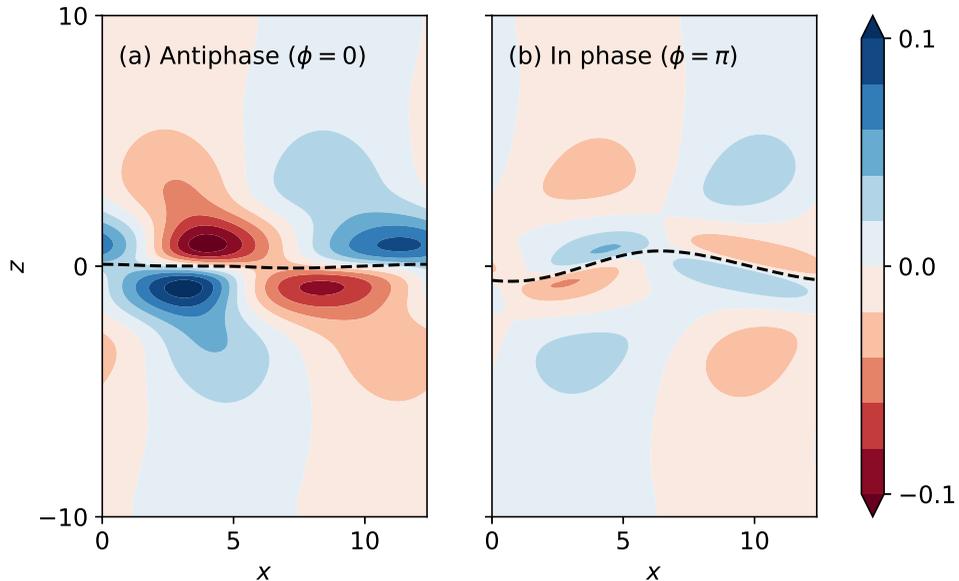}
	\centering\\
	\caption{Representative plots of the horizontal velocity perturbation, $u'$, for counter-propagating waves (\textit{a}) antiphase ($\phi = 0$, $t = 79$) and (\textit{b}) in phase ($\phi = \pi$, $t = 85$) in SWS. The solid line is the density interface. Note that the magnitude of $w'$ is similarly maximal at $\phi = 0$ and minimal at $\phi = \pi$.}
	\label{Singleledensity}
\end{figure}

\subsection{Single wavelength simulations}

\subsubsection{Counterpropagating Holmboe waves}
Counterpropagating Holmboe waves vary between an `antiphase' state ($\phi = 0$) and an `in phase' state ($\phi = \pi$), as in standing waves. Figure \ref{Singleledensity} is a plot of $u'$ when (a) $\phi = 0$ and (b) $\phi = \pi$. When the two waves are in antiphase ($\phi = 0$), the density interface is nearly horizontal, and the velocity perturbations are maxima in magnitude. While the two waves are in phase ($\phi = \pi$), the deflection of the density interface is the greatest and the corresponding velocity perturbations are minimal. The velocity and density perturbations do not vanish completely at any phase. 

\begin{figure}
	\includegraphics[scale=0.95]{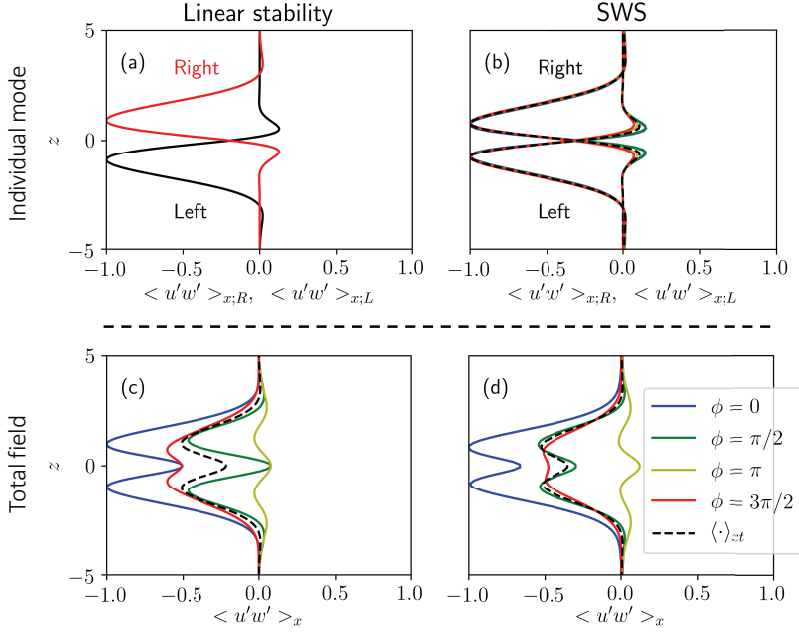}
	\centering\\
	\caption{Comparison of the normalised Reynolds stress between the linear stability analysis (the left column) and SWS (the right column) at different phases within a wave period. In linear stability analysis, four different phases with a $\pi/2$ interval are selected as representatives. The corresponding time in DNS are $t = 79$ ($\phi = 0$), $t = 82$ ($\phi = \pi/2$), $t = 85$ ($\phi = \pi$), and $t = 88$ ($\phi = 3\pi/2$) within a cycle. Above the dashed line, (\textit{a}) and (\textit{b}) are the Reynolds stresses of individual modes, i.e. rightward and leftward propagating modes. Below the dashed line, (\textit{c}) and (\textit{d}) are the total Reynolds stresses; $\langle \cdot \rangle_{xt}$ represents $\langle u'w' \rangle_{xt}$.}
	\label{Com_singlebothmodes}   
\end{figure}

\subsubsection{Comparison between DNS and linear stability analysis}
To compare the Reynolds stress in the DNS with linear theory, we separate the flow field associated with the rightward and leftward propagating modes (see Appendix \ref{appA}). Figure \ref{Linearanalysis} shows the evolution of horizontally averaged Reynolds stress $\langle u'w' \rangle_x$ in SWS for the total field in panel (e), the leftward propagating mode in panel (f), and the rightward propagating mode in panel (g). Note that the total Reynolds stress oscillates, and its vertical extend and magnitude increase with time. As the two waves propagate in opposite directions, the frequency of the oscillation in this pattern is double that of the individual modes ($2 \omega$; Ref. (\ref{PhasetermSimplifiedfinal})). This doubled frequency is consistent with the oscillation of the density field presented in figure \ref{Singleledensity}. 

Once the modes are separated with the Fourier transform, the periodicity disappears in both the leftward and rightward propagating modes (figure \ref{Linearanalysis}f and g); only a growing pattern remains. The magnitude is mainly concentrated above (below) the density interface for the rightward (leftward) propagating mode. It can be seen that the superposition of the two modes produces an additional phase dependent wave interaction field. This interaction field is shown in panel (h), obtained through the subtraction of (f)\&(g) from (e). A nearly symmetric pattern of growth and decay is observed in each half wave period. 

Figure \ref{Com_singlebothmodes}(a) \& (b) is a comparison of the mean vertical Reynolds stress profiles for each mode between linear stability analysis and SWS within one wave period ($t = 79-91$), normalised for direct comparison. Over one wave period in the SWS, the vertical expansion of the profiles is negligible for each mode. The peak value of Reynolds stress is located near $z$ = 1 ($z = -1$) for the rightward (leftward) propagating mode. The horizontally averaged Reynolds stress, $\langle u'w' \rangle_{x;R}$ or $\langle u'w' \rangle_{x;L}$, is generally independent of time except for small positive values around $z$ = $\pm$ 0.5 (panel b). This independence is further ideally confirmed in the linear stability analysis for an individual mode (panel a) illustrating that $\langle u'w' \rangle_x = \langle u'w' \rangle_t = \langle u'w' \rangle_{xt}$ (Ref. (\ref{singlemodeL})). 

The comparison of the total field between linear stability analysis and SWS is shown in figure \ref{Com_singlebothmodes} (c) and (d). As expected, the profiles vary with time. The maximum amplitude appears at $\phi = 0$ (`antiphase'), where instability extracts energy most efficiently from the mean flow. While positive value occurring at $\phi = \pi$ (`in phase') means that Holmboe instability returns energy back to the mean flow. On average over one period, the instability still extracts energy from the background mean flow, which is quite close to the amount at $\phi = \pi/2$ and $\phi = 3\pi/2$. The vertical structure of the Reynolds stress from the SWS agrees quite well with linear stability theory. 

\begin{figure}
	\includegraphics[scale=1]{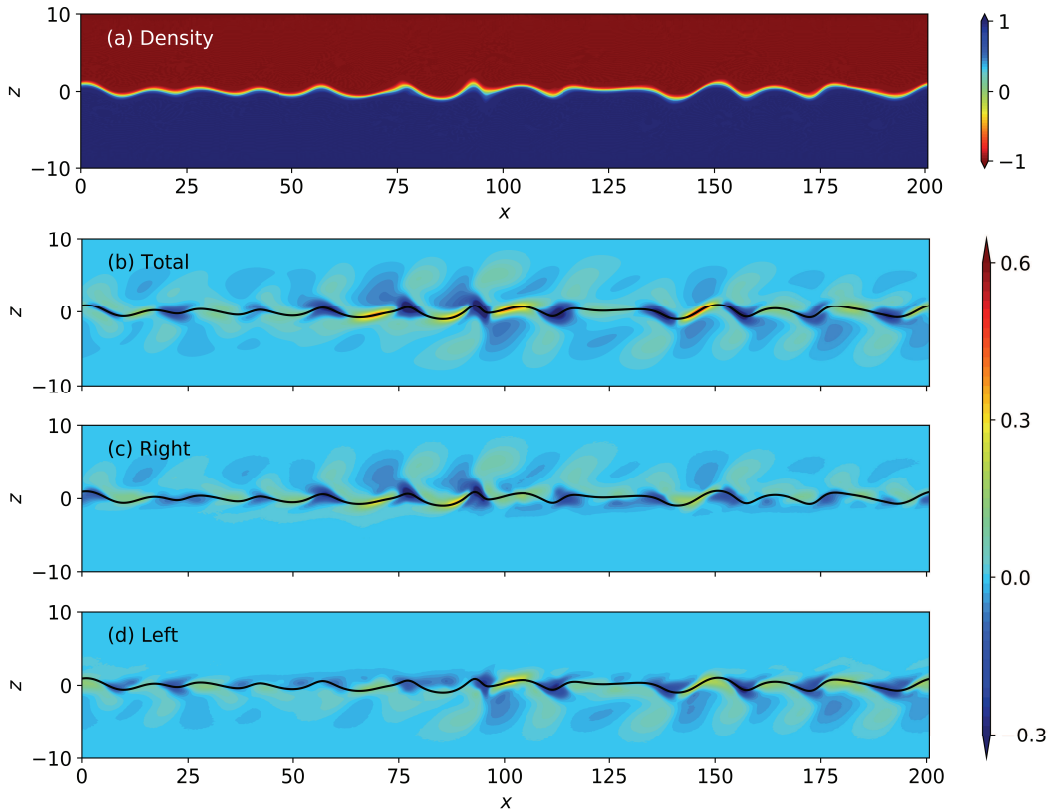}
	\centering\\
	\caption{Representative plots of the density fields (\textit{a}) and perturbation vorticity field (\textit{b})$-$(\textit{d}) at $t$ = 220 in MWS. (\textit{b}), (\textit{c}) and (\textit{d}) are the perturbation vorticity field for the total, rightward propagating, and leftward propagating wave modes, respectively. The black line represents the density interface.}
	\label{Multipledensity}  
\end{figure}

\subsection{Multiple wavelength simulations}
In the multiple wavelength simulations (MWS), we seeded a random initial perturbation, because seeding an eigenfunction perturbation in MWS, as we did in SWS, simply reproduces and copies the results of SWS over 16 wavelengths. 

\subsubsection{Wavenumber shifting}
The density and vorticity fields for MWS is shown in figure \ref{Multipledensity}. Starting with initial random perturbation at $t = $ 0, energy is extracted from the mean flow by the instability and fed into the wave field at, or very close to, the wavenumber of maximum growth, $k_{max}$. This results in approximately 16 wavelengths in the computational domain at early time. Evolving from the random noise, several waves with small amplitudes appear. As shown in figure \ref{Multipledensity}(a) for $t = $ 220, the amplitude of those waves grows and typical Holmboe wave cusps are observed pointing upward (propagating rightward) and downward (propagating leftward) directions. At this moment, the perturbation vorticity of the total, rightward, and leftward wave fields are shown in panels (b), (c), and (d) respectively. Similar to the direction of the cusp in the density field, the perturbation vorticities above (below) the density interface are associated with the rightward (leftward) propagating waves. Around 11 individual leading vorticities were observed at $t = 220$, representing the number of positive/ negative waves. Fewer than 16 waves indicates that some waves have merged (discussed later). From these separated perturbation vorticities (in panel c \& d), the waves are out of phase in $x-$direction showing different strength and shape in the leading vorticities. Due to the different relative phase between waves, it is not feasible to make a direct comparison of the horizontally averaged perturbation field as was shown in figure \ref {Com_singlebothmodes} for the SWS. 

\begin{figure}
	\includegraphics[scale=0.42]{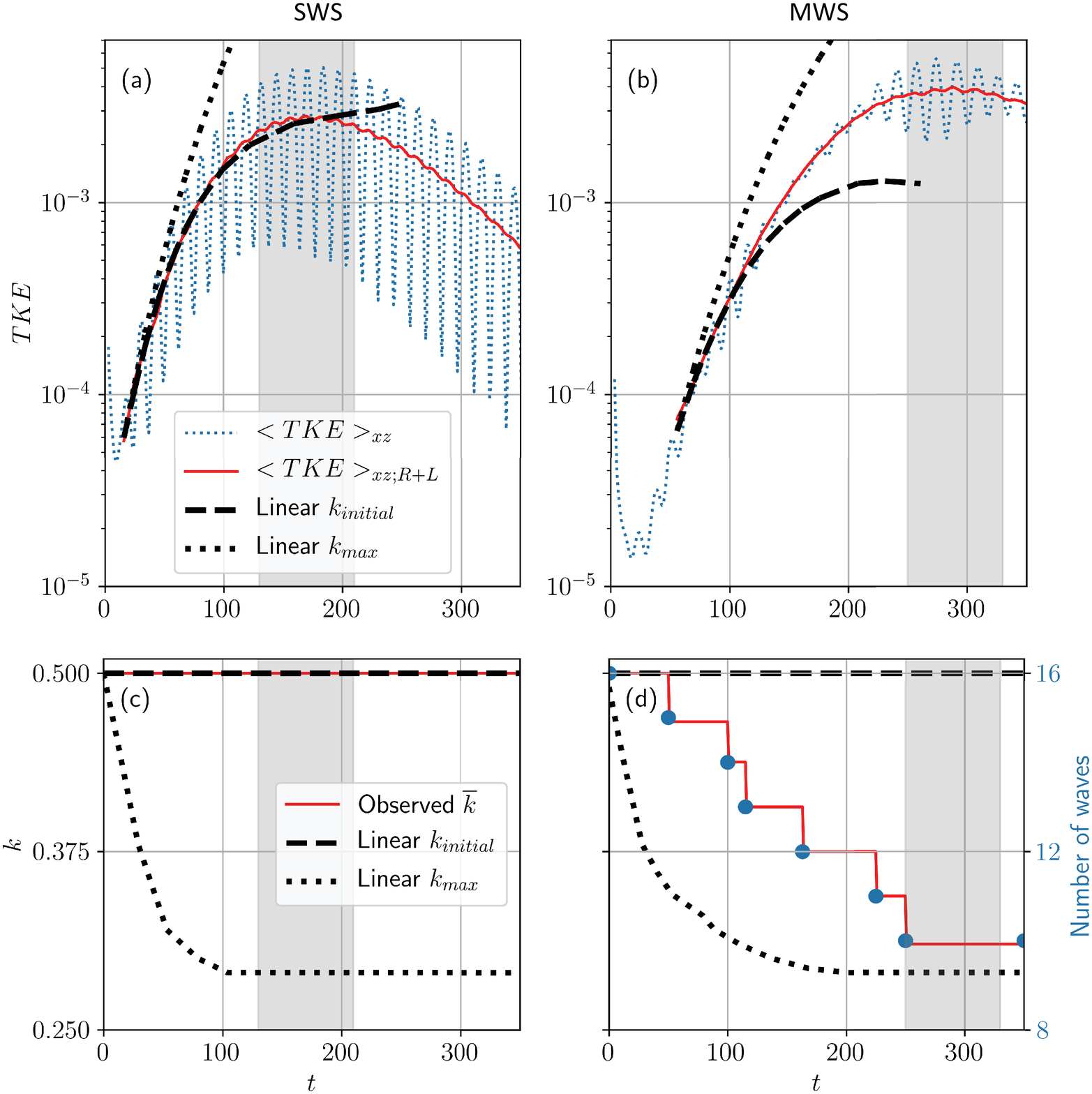}
	\centering\\
	\caption{Time evolution of the volume-averaged TKE and wavenumber in SWS (the left column) and in MWS (the right column): In (\textit{a},\textit{b}), the red solid line is the linear summation of the volume averaged TKE from the rightward and leftward propagating waves ($\langle TKE \rangle_{xz;R+L} = \langle TKE \rangle_{xz;R}+\langle TKE \rangle_{xz;L}$). The predicted growth rate from the linear stability analysis is denoted by the dashed black line (initial wavenumber $k_{initial} = 0.5$) and dotted black line ($k_{max}$), which is a function of time owing to the changing background profiles. In (\textit{c},\textit{d}), the black lines are from the same linear stability analysis and the red solid line is the observed average wavenumber in simulation with dots in (\textit{d}) representing the number of waves evolving from initial 16 waves ($k = $ 0.5) to 10 waves ($k = $ 0.31) in the MWS. The shaded period is selected as the stationary TKE period for analysis in \S 5.}
	\label{Growth}
\end{figure}

\subsubsection{Comparison of growth rate between SWS and MWS}
The evolution of TKE for SWS and MWS is plotted on a log scale in figure \ref{Growth}. After a start-up period in which energy of the initial perturbation rapidly decays, the waves grow. In the case of the SWS (figure \ref{Growth}a), the volume-averaged TKE ($\langle TKE \rangle_{xz}$) shows a strong oscillation throughout the whole simulation. This oscillation has a period of $2\omega$ and is a result of energy exchange between the perturbation kinetic energy and perturbation potential energy \citep[approximately $\propto \rho'^2$;][]{kaminski2014transient} in the standing wave highlighted in figure \ref{Singleledensity}. Once the leftward and rightward modes are separated and their individual kinetic energies linearly summed ($\langle TKE \rangle_{xz;R+L} = \langle TKE \rangle_{xz;R}+\langle TKE \rangle_{xz;L}$), the oscillation is absent. The TKE for individual modes is approximately equal, $i.e.$  $\langle TKE \rangle_{xz;R} \approx \langle TKE \rangle_{xz;L}$. Accounting for the evolving background profiles, the TKE estimated from linear stability theory is also shown in figure\ref{Growth}(a). The growth rates of the initial wavenumber $k_{initial} = 0.5$ and the wavenumber of maximum growth $k_{max}$ are shown by the dashed and dotted black lines, respectively. These growth rates are calculated at each time step using the evolving velocity and density profiles. Both the growth rates continuously decrease due to the diffusion of the background flow over time. In the SWS, the growth rate of $k_{max}$ overestimates the TKE, while that of $k_{initial} = 0.5$ successfully predicts the TKE during the growth period since the wavenumber is fixed ($k = k_{initial}$ in figure \ref{Growth}c). Once the waves saturate ($t>160$), the linear prediction ($k_{initial}$) begins to deviate from the SWS indicating the dominance of nonlinear processes. During this nonlinear period, the separation of the TKE for the rightward and leftward propagating waves remains effective.  

Unlike the strong oscillation in the volume-averaged TKE in the SWS, the growth of TKE in MWS is relatively steady (figure \ref{Growth}b). This is because the leftward and rightward propagating waves have a distribution of phases and amplitudes in $x$ at any given time ($e.g.$ figure \ref{Multipledensity}) resulting in less coherent interferences. Thus, the volume-averaged TKE is close to the linear summation of the rightward and leftward waves ($\langle TKE \rangle_{xz}~ v.s.~ \langle TKE \rangle_{xz;R+L}$) when $t \gtrsim 55$. Before $t=55$, the two wave modes cannot be accurately separated by the Fourier transform due to the non-modal Holmboe instability \citep{guha2014wave}. Even during the modal development period ($t \gtrsim 55$), the growth rate of TKE in the MWS cannot be straightforwardly compared with the linear stability analysis as the number of Holmboe waves (and thus the relevant choice of wavenumber) is evolving in time. The predicted growth rate from the linear stability analysis with the initial wavenumber ($k_{initial} = 0.5$) and the wavenumber of maximum growth ($k_{max}$) is also shown in figure \ref{Growth}(b). The TKE in the MWS evolves between these two predicted results. The wave merging in the MWS results in a maximum TKE that is approximately twice that in the SWS. Additional simulations (not included) with horizontal domain lengths between $L_x = \lambda_0$ and $L_x = 32\lambda_0$ indicated that a domain longer than $16\lambda_0$ was found to have a negligible influence on the saturation TKE.     

The wave merging events are similar to those reported in \citet{carpenter2010holmboe}. This process of losing waves results in an observed wavenumber that is continually shifted downwards. The merging events are reported to undergo a vortex pairing process similar to those found in simulations of KH instabilities \citep[e.g.][]{patnaik1976numerical, dong2019sensitivity, guha2019predicting} and observed in Holmboe instability through experiments in \citet{lawrence1991stability}.  

The decreasing wavenumber of instabilities in the MWS is plotted in figure \ref{Growth}(d). The wavenumber of maximum growth rate, $k_{max}$, from linear stability theory is plotted as the dashed black line. Limited by the periodic boundary conditions, the observed wavelength evolves in discrete steps. In contrast, the linear stability theory is not limited to steps as it is not constrained in the horizontal domain. Unlike the fixed wavenumber in the SWS (figure \ref{Growth}c), the decrease of $k_{max}$ in time is due to the increasing shear layer thickness that results from diffusion and then presents in the Holmboe waves. The growing mode in the MWS at every time step is not necessarily the `instantaneously' most unstable mode in linear stability analysis.  The observed wavenumber is distributed between $k_{initial}$ and $k_{max}$. Since the growth rate of the instability does not rely on one certain wavenumber, its associated TKE is distributed across $k_{initial}$ (16 waves) to $k = 0.31$ (final; 10 waves) over time. However, both the observed average wavenumber ($\overline{k}$) and estimation ($k_{max}$) approach a similar value of $k = $ 0.31 for $t > $250. During this later period, the TKE is also relatively steady; this approximate stationary period is to be used in the next section.

\section{Reynolds stress comparisons}
\begin{figure}
	\includegraphics[scale=0.38]{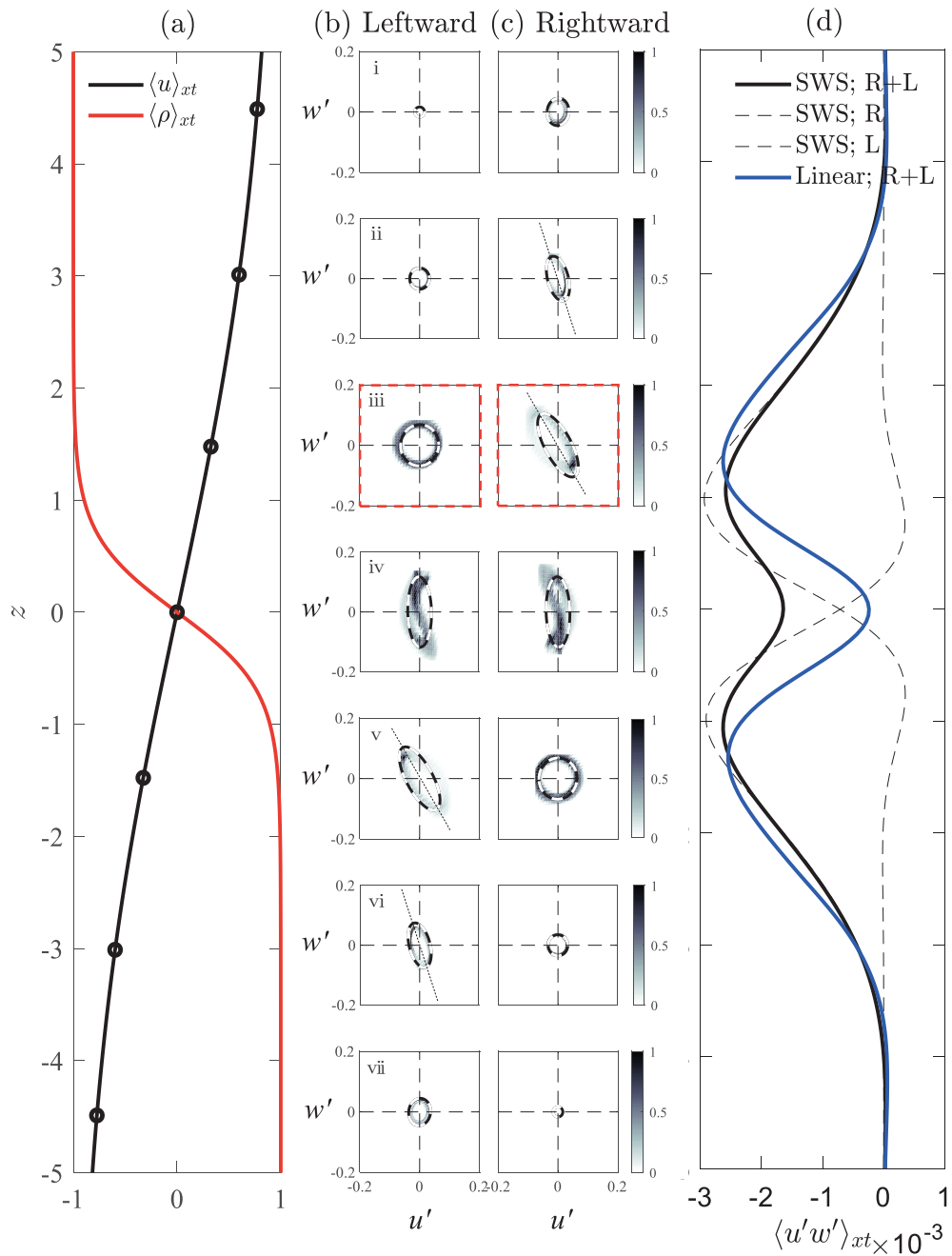}
	\centering\\
	\caption{Flow fields of SWS for the period of stationarity ($t = 130 - 210$) and linear stability analysis ($k = 0.5$). (\textit{a}) mean density and velocity profiles from the SWS. The black circles indicate the elevations of ($u',w'$)-pairs in (\textit{b}) and (\textit{c}) with a spacing of 1.5. (\textit{b}) and (\textit{c}) the comparison of ($u',w'$)-pairs between the linear stability analysis and the SWS for the leftward and rightward propagating modes, respectively. In the SWS, the ($u',w'$)-pairs are presented based on joint probability density functions (PDFs) where the darker color denotes the higher probability. The ellipses in the black-white line is from the linear stability analysis and its magnitude is scaled with the SWS; their major axes are illustrated by the black dashed lines. (\textit{d}) is the Reynolds stress. The ($u',w'$)-pairs tilt towards the 2nd \& 4th quadrants within the shear layer below (above) the density interface for the leftward (rightward) mode. In (\textit{b}) and (\textit{c}), the superposition of ($u',w'$)-pairs with the red boxes (iii. at $z =$ 1.5) is analysed in figure \ref{DiscussionMWS2}.}
	\label{DiscussionSWS}
\end{figure}

In this section, we compare ($u',w'$)-pairs predicted using linear stability theory with the probability density functions (PDFs) of ($u',w'$) obtained using DNS, and those measured in the laboratory experiments of \citet{lefauve2019regime}. The resultant Reynolds stresses are also compared. The linear stability analysis used the appropriate mean density and velocity profiles: i.e., the profiles obtained during the periods of approximate stationarity in the DNS (see figure \ref{DiscussionSWS}a), and the profiles obtained during the period of quasi-steady flow in experiment H3 of \citet[]{lefauve2019regime, lefauve2019research}

\subsection{Comparison with DNS}
The ($u',w'$)-pairs for the leftward and rightward propagating waves, obtained from linear stability analysis, are compared with the SWS during the period of stationarity from $t = 130 - 210$, in figure \ref{DiscussionSWS}(b) \& (c).  In the linear stability analysis the ($u',w'$)-pairs form either circles or ellipses. Within the shear layer below the density interface, the ($u',w'$)-pairs associated with the leftward propagating waves form ellipses that are oriented towards the 2nd and 4th quadrants (figure \ref{DiscussionSWS}b), and the corresponding Reynolds stresses are negative (figure \ref{DiscussionSWS}d). Above the density interface, the ($u',w'$)-pairs associated with the leftward propagating waves form circles, and do not contribute to the Reynolds stress. The ($u',w'$)-pairs for the rightward propagating waves mirror those of the leftward propagating waves (figure \ref{DiscussionSWS}c), as do their Reynolds stresses (figure \ref{DiscussionSWS}d).  The joint PDFs of the ($u',w'$)-pairs obtained in the SWS appear as `doughnut'-shaped clouds that, in general, closely match the ellipses and circles predicted using linear stability analysis (figure \ref{DiscussionSWS}b \& c). However, the comparison is not as good near to $z = 0$, due to nonlinear interactions between the leftward and rightward propagating waves (panel iv in figure \ref{DiscussionSWS}b \& c).

The vertical profiles of Reynolds stress have the same basic shape in both the SWS and the linear stability analysis (figure \ref{DiscussionSWS}d). The profiles are symmetric about $z = 0$, with a minimum at $z = 0$ and two peaks at $z \approx \pm 1.3$, beyond which the Reynolds stresses decay to zero at the edge of the shear layer. The peak above (below) the interface is due to the rightward (leftward) propagating wave. The largest difference between the simulations and linear stability analysis occurs again at $z = 0$, where the nonlinear interactions between the counterpropagating waves is the greatest, again due to vertical oscillations of the density interface in the SWS (figure \ref{Singleledensity}).

\begin{figure}
	\includegraphics[scale=0.76]{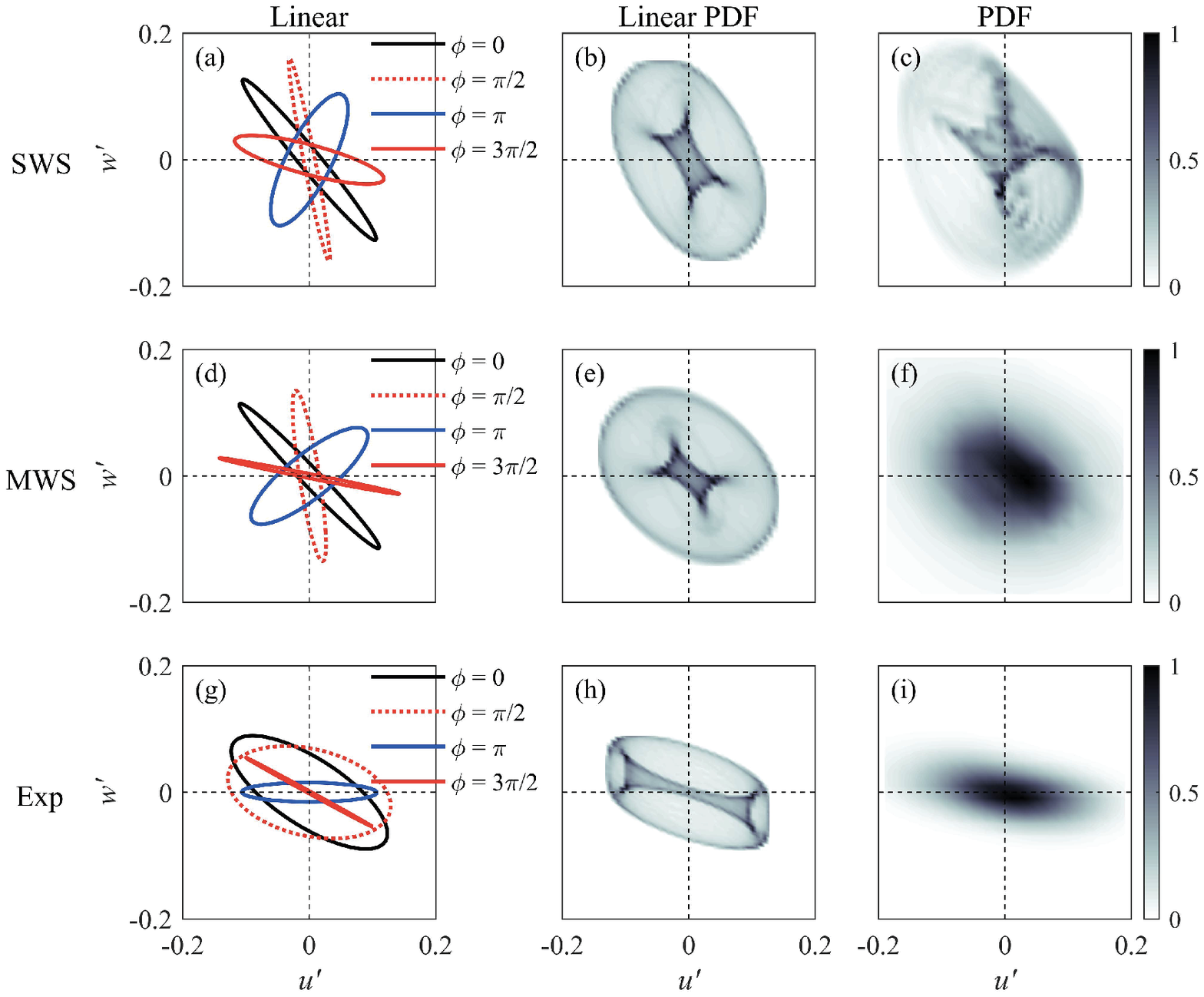}
	\centering\\
	\caption{Anisotropic perturbation fields in the superposition of the rightward and leftward propagating waves: the complexity increases from SWS (the upper row) to MWS (the middle row), and further to laboratory experiment (the lower row). The left column is ($u',w'$)-pairs at four different phases ($\phi$ = 2$\omega t$) with a $\pi/2$ interval in the linear stability analysis. The middle column corresponds to the ($u',w'$) presented based on joint PDFs. The right column shows the joint PDFs of ($u',w'$) around the vertical level of the peak Reynolds stress in (\textit{c}) SWS (the red box at $z = 1.5$ in figure \ref{DiscussionSWS}), (\textit{f}) MWS, and (\textit{i}) laboratory experiment of \citet{lefauve2019regime}. Multiple wavenumbers exist in the flow of MWS and experiments such that ($u',w'$)-pairs show solid-cloud structures.}
	\label{DiscussionMWS2}
\end{figure}

To further compare linear stability predictions with the SWS, and investigate the impact of the counter-propagating waves, we examine the ($u',w'$)-pairs resulting from the superposition of the leftward and rightward propagating waves at $z = 1.5$, which is close to the vertical location of the peak Reynolds stress (figure \ref{DiscussionMWS2}). In figure \ref{DiscussionSWS}(b) \& (c) we presented the ($u',w'$)-pairs predicted by linear stability theory separately for the leftward and rightward propagating waves; when these ($u',w'$)-pairs are added together, the orientation and aspect ratio of the resulting ellipses are phase dependent, as shown in figure \ref{DiscussionMWS2}(a).  At $\phi = 0, \pi/2$, and $3\pi/2$, the major axes of the ($u',w'$)-ellipses tilt toward the 2nd \& 4th quadrants, resulting in negative Reynolds stresses ($\langle u'w' \rangle_x < 0$); while at $\phi = \pi$, the ($u',w'$)-pairs tilt toward the 1st \& 3rd quadrants, resulting in positive Reynolds stresses ($\langle u'w' \rangle_x > 0$). Similar results for the temporal evolution of $\langle u'w' \rangle_x$ during the linear growth are shown in figures \ref{Linearanalysis}(a) and \ref{Com_singlebothmodes}(c).

When the ($u',w'$)-pairs of the counterpropagating waves are combined over a full wave period, the joint PDF of ($u',w'$) is topologically similar to a four-spoked `steering wheel' (figure \ref{DiscussionMWS2}b). The vertices (along major axes) of these ($u',w'$)-ellipses combine to form the outer rim of the steering wheel; whereas, the co-vertices (along minor axes) combine to form the spokes of the steering wheel.  The corresponding PDF of the ($u',w'$)-pairs from the SWS is shown in figure \ref{DiscussionMWS2}(c). While somewhat distorted, the tilted elliptical shape with steering wheel features, is still apparent and similar to the linear stability predictions. The differences between them are presumably due to the slow growth and decay of instabilities during the period of approximate stationarity.

The linear stability predictions from the MWS are very similar to those of the SWS (figure \ref{DiscussionMWS2}d \& e) with minor differences attributable to the slightly different mean velocity and density profiles during their periods of stationarity. However, the PDF of ($u',w'$) forms a unimodal cloud in the MWS, rather than the steering wheel pattern found in the SWS (figure \ref{DiscussionMWS2}c) and the linear stability prediction (figure \ref{DiscussionMWS2}e). This unimodal cloud reflects the complicated interactions between multiple counterpropagating waves of varying phase, amplitude, and wavelength in the MWS (figure \ref{DiscussionMWS2}f). Nevertheless, the cloud has the same orientation as the steering wheel pattern predicted with linear stability theory.  

\subsection{Comparison with laboratory experiment}
The results thus far have progressed in complexity from linear stability analysis, to single wave length DNS, and to multiple wave length DNS. Now we compare our results with laboratory measurements of Holmboe instabilities in a two-layer exchange flow investigated by \citet[Exp. H3]{lefauve2019regime}. The parameters of the laboratory experiment were different from those of the SWS and MWS, but sufficiently similar to warrant the qualitative comparisons that we make here. 

\begin{figure}
	\includegraphics[scale=0.7]{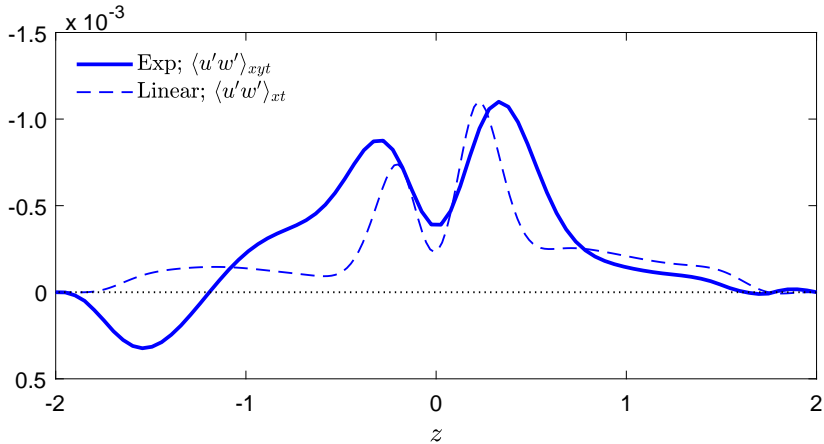}
	\centering\\
	\caption{Vertical profiles of the total Reynolds stress in the laboratory experiment and the linear stability analysis ($k = $ 1.2).}
	\label{DiscussionH3}
\end{figure}

We follow the same procedure with respect to comparing linear stability theory with the laboratory experiments as we did with the SWS and MWS. Qualitatively, the linear stability predictions for the laboratory experiments are the same as those for the SWS and MWS (figure \ref{DiscussionMWS2}g \& h). However, the orientation and aspect ratio of the ($u',w'$) ellipses, and the steering wheel structure, are slightly different. This difference is not surprising since the background mean velocity profiles in the laboratory experiment are nearly sinusoidal due to the no slip condition at the top and bottom channel walls, as opposed to the hyperbolic tangent profiles in the SWS and MWS. The joint PDF of ($u',w'$) in the laboratory experiment is similar to the unimodal structure of the MWS, due to the coexistence of multiple Holmboe wavelengths (figure \ref{DiscussionMWS2}i). This PDF cloud resembles those characteristic of fully turbulent fields \citep{tennekes1972first}. It also shows a similar orientation as that predicted from linear stability theory.

The vertical profile of the total Reynolds stress in the laboratory experiment is similar to that predicted by the linear stability analysis (figure \ref{DiscussionH3}). As in the SWS and MWS, two peaks appear in the vertical profile, one above and one below the density interface.  A slight asymmetry in the Reynolds stress profiles for the laboratory experiment results from the experimental configuration, which leads to a slightly stronger vorticity above the interface than below the interface \citep[]{lefauve2018structure, lefauve2019regime}. There is also a region of positive Reynolds stress below the interface in the laboratory experiment, which again is likely due to the irregular nature of the velocity profile.

\section{Summary and conclusions} 
In this paper, we investigated the Reynolds stress generated by Holmboe instabilities in stratified shear flows. Linear stability analysis was used to explain the generation of Reynolds stresses. Then, single wavelength simulations (SWS) and multiple wavelength simulations (MWS) were used to study the effects of wavenumber shifting on the instabilities from linear growth to saturation, and the influence of the presence of multiple wavelengths on Reynolds stresses. Finally, we analysed the Reynolds stresses in a relevant laboratory experiment.

Counterpropagating symmetric Holmboe waves form an oscillation in the flow perturbation fields. The corresponding Reynolds stresses depend on the phase of counterpropagating waves. This oscillation is separated into the leftward and rightward propagating components with the discrete Fourier transform, enabling a direct comparison of the perturbation field and its corresponding Reynolds stress between simulations and linear stability analysis. 

Linear stability analysis predicts that the ($u',w'$)-pairs associated with the leftward propagating waves form ellipses within the shear layer below the density interface. These ellipses are orientated towards the 2nd \& 4th quadrants. The ($u',w'$)-pairs for the rightward propagating waves mirror those of the leftward propagating waves. These ($u',w'$)-ellipses are generated by the elliptical trajectories of partical orbits in Holmboe waves. When combining the leftward and rightward modes, ($u',w'$)-pairs are also ellipses whose orientation and aspect ratio are phase dependent. The corresponding joint PDFs of ($u',w'$) over a full wave period have a `steering wheel' structure. 

In the SWS, joint PDFs of ($u',w'$) for either the rightward or the leftward propagating waves are doughnut shaped ellipses, which closely match the ellipses predicted using linear stability analysis. Combining the leftward and rightward modes, joint PDFs of ($u',w'$) yield a `steering wheel' structure similar to the linear theory predictions. In the MWS and laboratory experiments, the presence of multipe waves with varying wavelengths, phases, and amplitudes smears out the `steering wheel' structure, leaving an elliptical cloud with similar orientation to the corresponding linear prediction.

The linear stability analysis and statistical analysis of the DNS and laboratory experiments yield similar results: the ($u',w'$)-pairs predominantly oriented towards the 2nd \& 4th quadrants resulting in negative Reynolds stresses. The vertical structure of the Reynolds stress exhibits two peaks, above and below the density interface. The upper and lower peaks are respectively caused by the rightward and leftward propagating Holmboe waves.

~~
\newline
\newline
This study is supported by the Natural Sciences and Engineering Research Council of Canada.

\appendix
\section{Separation of rightward and leftward propagating waves in simulations}\label{appA}

\begin{figure}
	\includegraphics[scale=0.65]{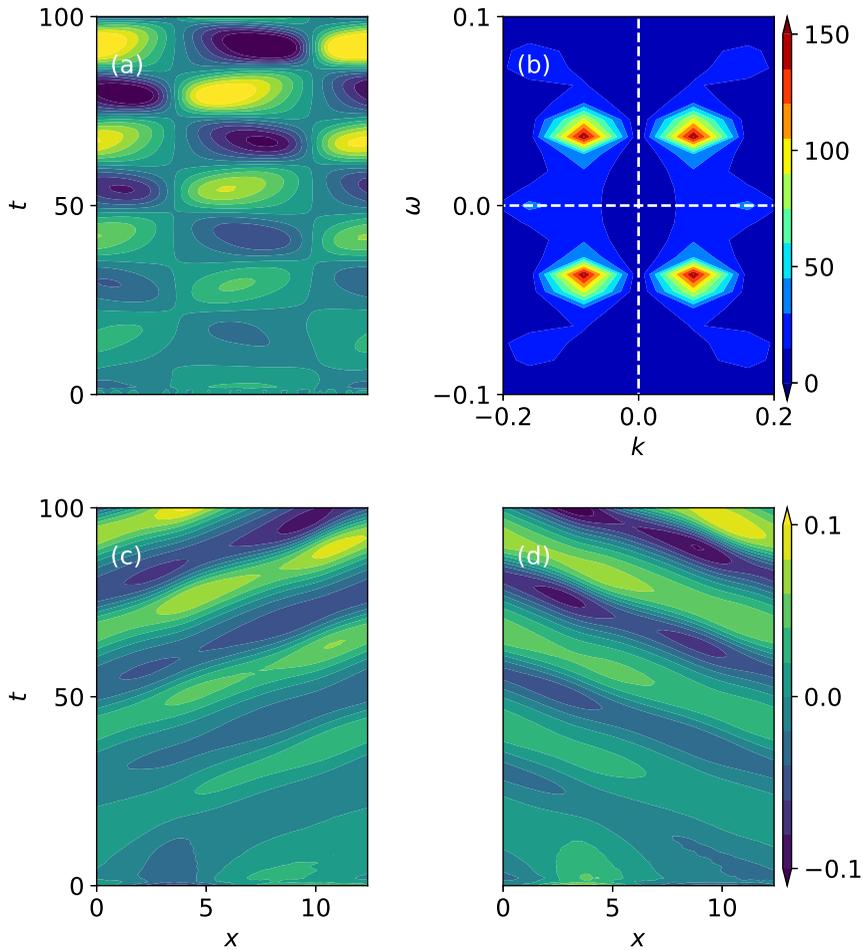}
	\centering\\
	\caption{$w'$ at $z = 0$ separation for rightward and leftward propagating waves in SWS: (\textit{a}) vertical velocity perturbation characteristics, (\textit{b}) Dispersion relation is obtained by two-dimentional (2D) Fourier transform from characteristics (\textit{a}). $w'$ for (\textit{c}) the rightward propagating and (\textit{d}) leftward propagating waves are obtained by 2D inverse Fourier transform from the dispersion relation based on the energy in the $k \omega > 0 $ and $k \omega < 0$ quadrants of panel (\textit{b}), respectively.}
	\label{Technique}
\end{figure}
Figure \ref{Technique}(a) is a plot of the vertical velocity perturbation at the mid depth compiled over time. Although the velocity consists of contributions from both rightward and leftward propagating waves, we have filtered the characteristics to reveal only the rightward and leftward propagating wave modes. This technique has been used in the separation of the wave characteristics \citep[e.g.][]{tedford2009symmetric, carpenter2010holmboe}. We perform a two-dimensional Fourier transform of the $w'$ resulting in a wavenumber-frequency ($k-\omega$) spectrum, whose amplitude is shown in figure \ref{Technique}(b). The rightward propagating waves are then removed by setting the quadrants (the complex value) in which $k$ and $w$ have the same sign (1st \& 3rd quadrants) to zero and performing the inverse transform. The same procedure is applied to the opposite quadrants (2nd \& 4th quadrants) to remove the leftward propagating waves. The $w'$ for the rightward and leftward propagating waves are presented in figure \ref{Technique}(c) and (d), respectively. This pattern is highly similar to wave characteristics. The vertical velocity perturbation can be approximately formulated as $w' = \frac{D \eta}{Dt}$ and thus only a phase difference occurs between wave characteristics and $w'$ characteristics. The dispersion relationship in both $w'$ and wave characteristics shows the same shape and amount of energy in positive and negative wave modes. By taking $d\omega / d k$ in the dispersion relationship, the rightward and leftward waves yield the same phase speed propagating in opposite directions, as indicated by traditional linear stability analysis \citep[e.g.][]{holmboe1962behavior, lawrence1991stability}.

\begin{figure}
	\includegraphics[scale=0.65]{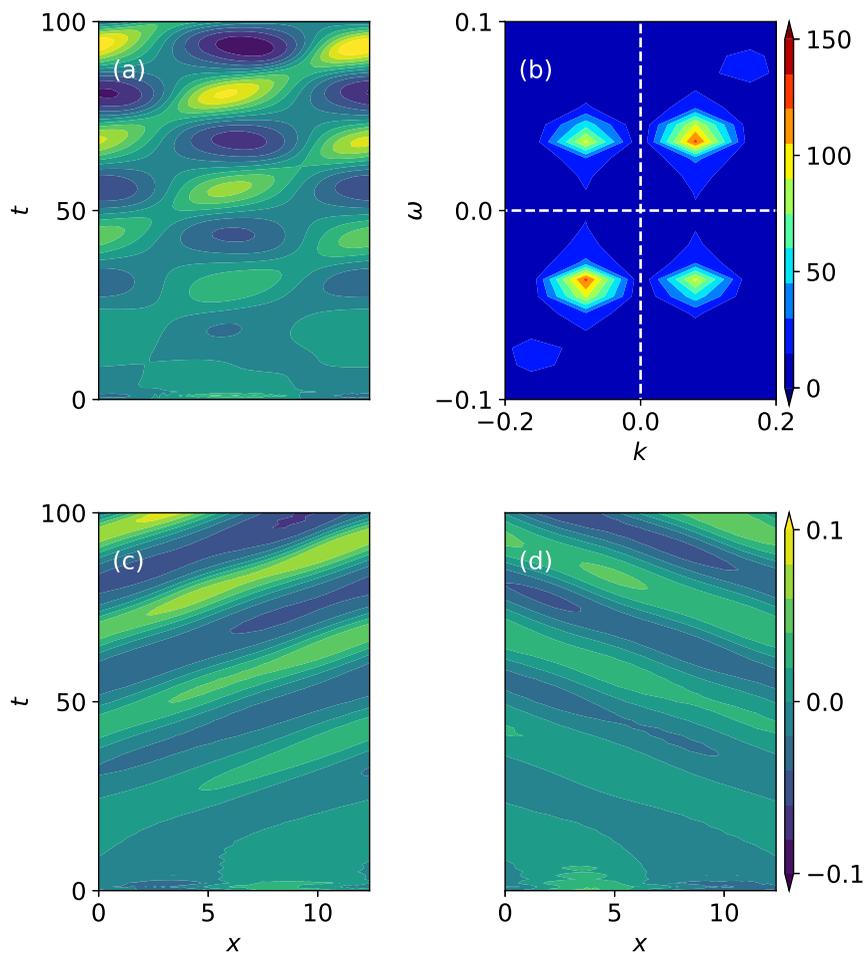}
	\centering\\
	\caption{$w'$ at $z = 1.5$ separation for rightward and leftward propagating waves in SWS. See figure \ref{Technique} for details.}
	\label{Technique3}
\end{figure}
This separation method was used for $w'$ and $u'$ for every $z$ location, and thus the whole velocity perturbation field was separated for the associated rightward and leftward propagating wave modes. The $w'$ separation of another location, for example $z = 1.5$, is shown in figure \ref{Technique3}. The magnitude of $w'$ is smaller than that at $z = 0$ as shown in figure \ref{Technique}. In figure \ref{Technique3}(b), the energies in spectrum for the rightward and leftward propagating waves are presented in the 1st \& 3rd, and 2nd \& 4th quadrants, respectively. The energy of the rightward propagating mode is larger than that of the negative propagating mode, as with any loctions above the density interface. However, below the density interface, the energy of leftward propagating mode dominates. The separated perturbation velocity field is then used for the Reynolds stress and TKE.

\clearpage
\bibliographystyle{jfm}
\bibliography{jfm}

\clearpage
\tableofcontents

\end{document}